\documentclass[%
preprint,
superscriptaddress,
amsmath,amssymb,
aps,
prb,
]{revtex4-2}

\usepackage[english]{babel}
\usepackage[utf8]{inputenc}
\usepackage{amsthm}
\usepackage{mathtools}
\usepackage{physics}
\usepackage{xcolor}
\usepackage{graphicx}
\usepackage[left=23mm,right=13mm,top=35mm,columnsep=15pt]{geometry} 
\usepackage{adjustbox}
\usepackage{placeins}
\usepackage[T1]{fontenc}
\usepackage{lipsum}
\usepackage{csquotes}\usepackage[pdftex, pdftitle={Article}, pdfauthor={Author}]{hyperref} 
\usepackage{isotope}
\begin{document}
	\title{Formation of Core-Shell Precipitates in off-stochiometric Ni-Mn-Sn Heusler alloys probed through the induced Sn-moment}
	
	\author{Benedikt Eggert}
	\email[Correspondence email address: ]{Benedikt.Eggert@uni-due.de}
	\affiliation{Faculty of Physics and Center for Nanointegration Duisburg-Essen (CENIDE), University of Duisburg-Essen, Lotharstr.1, 47057 Duisburg, Germany}
	\author{Asl\i\,\c{C}ak\i r}
	\affiliation{Department of Metallurgical and Materials Engineering, Mu\u{g}la Sıtkı Ko\c{c}man University, 48000 Mugla, Turkey}
	\author{Damian Günzing}
	\affiliation{Faculty of Physics and Center for Nanointegration Duisburg-Essen (CENIDE), University of Duisburg-Essen, Lotharstr.1, 47057 Duisburg, Germany}
	\author{Nicolas Josten}
	\affiliation{Faculty of Physics and Center for Nanointegration Duisburg-Essen (CENIDE), University of Duisburg-Essen, Lotharstr.1, 47057 Duisburg, Germany}
	\author{Franziska Scheibel}
	\affiliation{Functional Materials, Institute of Materials Science, Technische Universität, 64287 Darmstadt, Germany}
	\author{Richard A. Brand}
	\affiliation{Faculty of Physics and Center for Nanointegration Duisburg-Essen (CENIDE), University of Duisburg-Essen, Lotharstr.1, 47057 Duisburg, Germany}
	\author{Michael Farle}
	\affiliation{Faculty of Physics and Center for Nanointegration Duisburg-Essen (CENIDE), University of Duisburg-Essen, Lotharstr.1, 47057 Duisburg, Germany}
	\author{Mehmet Acet}
	\affiliation{Faculty of Physics and Center for Nanointegration Duisburg-Essen (CENIDE), University of Duisburg-Essen, Lotharstr.1, 47057 Duisburg, Germany}
	\author{Heiko Wende}
	\affiliation{Faculty of Physics and Center for Nanointegration Duisburg-Essen (CENIDE), University of Duisburg-Essen, Lotharstr.1, 47057 Duisburg, Germany}
	\author{Katharina Ollefs}
	\affiliation{Faculty of Physics and Center for Nanointegration Duisburg-Essen (CENIDE), University of Duisburg-Essen, Lotharstr.1, 47057 Duisburg, Germany}
	
	\date{\today} 
	
	\begin{abstract}
		\noindent The Shell-ferromagnetic effect originates from the segregation process in off-stochiometric Ni-Mn-based Heusler. In this work, we investigate the precipitation process of L2$_1$-ordered Ni$_2$MnSn and L1$_0$-ordered NiMn in off-stochiometric Ni$_{50}$Mn$_{45}$Sn$_{5}$ during temper annealing, by X-ray diffraction (XRD) and \isotope[119]Sn M\"ossbauer spectroscopy. While XRD probes long-range ordering of the lattice structure, Mössbauer spectroscopy probes nearest-neighbour interactions, reflected in the induced Sn magnetic moment. As shown in this work, the induced magnetic Sn moment can be used as a detector for microscopic structural changes and is, therefore, a powerful tool for investigating the formation of nano-precipitates. Similar research can be performed in the future, for example, on different pinning type magnets like Sm-Co or Nd-Fe-B.
	\end{abstract}

	\maketitle
	
	\section{Introduction}
	
	Due to the multifaceted phase diagram~\cite{akr2013,akr2015} of magnetic Heusler alloys, this material class possesses a variety of interesting phenomena~\cite{GRAF20111}. For example, NiMn-Heusler alloys show a magnetostructural phase transition~\cite{Ollefs2015ShapeMemory} or intrinsic exchange bias~\cite{Cakir2016} due to the presence of mixed magnetic interactions (antiferromagnetic and ferromagnetic)~\cite{aksoy2009,entel2014, monroe2015,entel2018,sokolov2019}. Besides, it shows the possibility of adjusting the magnetic anisotropy energy, e.g. by interstitial doping~\cite{Gao2020} and, therefore, serves as a prototype system for investigations of fundamental physical phenomena such as structural disorder~\cite{Schleicher2017}. These properties make it possible that Heusler alloys can be used for applications, for example, in the area of magnetic sha\-pe-mem\-ory~\cite{Bachaga2019}, magnetocalorics~\cite{Gottschall2018} and spintronics~\cite{Karel2017}. 
	Heusler alloys have a huge potential for applications, however, off-stoichiometric variations of Heusler alloys suffer due to a tendency of segregation. Sokolovskiy et al.~\citenum{sokolov2019} recently performed DFT-calculations and showed that off-stochiometric Mn-rich Ni$_2$Mn$_{1+x}$(In,Sn,Al)$_{1-x}$ compounds are unstable at low temperatures and decompose into a dual phase system. However, it is possible to utilize this process, as it will be discussed in the following.
	The shell-ferromagnetic effect is a newly achieved property of the off-stoichiometric Heusler compound, which is less well studied and occurs in Mn-rich antiferromagnetic (AF) Heusler-based compounds~\cite{Krenke2016} and opens paths to different functionalities. This effect occurs when $\text{Ni}_{50}\text{Mn}_{45}\text{Z}_{5}$ (Z: Al, Ga, In, Sn, Sb) decomposes into cubic L$2_1$ ferromagnetic (FM) Heusler Ni$_{50}$Mn$_{25}$Z$_{25}$ and L$1_0$-ordered AF Ni$_{50}$Mn$_{50}$ during temper-an\-neal\-ing at temperatures around $600\,\mathrm{K}$ $< T_A <$ $750 \,\mathrm{K}$, where $T_A$ is the annealing temperature. By applying a magnetic field during the annealing process, nano precipitates are formed within a strongly pinning AF matrix, originating from Ni-Mn and off-stochiometric Ni$_{50}$Mn$_{45}$Z$_{5}$. A collection of these nano precipitates in a macroscopic sample gives rise to a partially compensated magnetic response to an applied magnetic field, which has been demonstrated in a video~\cite{YoutubeShellFM1}. The observed pinning mechanism implies that the formed precipitates could form building blocks for high-per\-for\-mance and lightweight permanent magnets of unsurpassed coercivity. The magnetic moment of Ni-Mn, becomes pinned in the direction of the applied magnetic field during annealing so that the field-dependence up to 9\,T appears as a vertically shifted hysteresis loop, while it is a minor loop within a major loop with a coercive field exceeding 5\,T~\cite{Scheibel2017Room-temperatureShell-ferromagnet}. The remanent magnetisation of the loop is always positive and only reorientates entirely in fields exceeding 20\,T ($T < 550\mathrm{\,K}$). The core structure of the precipitate (Ni$_2$MnZ) is, however, magnetically soft, and the spins rotate freely in the direction of an applied magnetic field. These structures were first found as a result of decomposing Ni$_{50}$Mn$_{45}$In$_{5}$~\cite{akr2016} or Ni$_{50}$Mn$_{45}$Ga$_{5}$~\cite{Krenke2016} at 650\,K in a magnetic field. For the effective compensation of the magnetisation, the surface-to-volume ratio of the precipitate is important. For the case of a large surface-to-volume ratio, the magnetisation of the Ni$_2$MnZ-cluster can be compensated by the Ni-Mn surrounding. For larger Ni$_2$MnZ-precipitates, the Ni-Mn-surrounding is not sufficient for the compensation of Ni$_2$MnZ spins, and the shell-ferromagnetic effect does not occur. Scherrer analysis indicates a precipitate size of 3-5\,nm for Ni$_{50}$Mn$_{45}$In$_{5}$ annealed at 650 K, corresponding to a surface-to-volume ratio~\cite{Dincklage2018} of 1.2–2\,nm$^{-1}$, while for Ni$_{50}$Mn$_{45}$Sb$_{5}$ the precipitate size is in a similar range of 5-10\,nm~\cite{Wanjiku2019}. Besides the potential use case in permanent magnets, these magnetically pinned precipitates can be used in materials possessing a first-order magnetostructural phase transition~\cite{akr2020}, where the precipitates may serve as a nucleation site for the phase transition. In this case, the precipitates may induce a local strain field that can energetically favour the martensite-austenite transition. This mechanism has the potential to improve magnetocaloric properties~\cite{amaz2020} of this compound or can increase the mechanical stability~\cite{Pfeuffer2021}.
	
	Within the current work, we report on our recent findings characterising Ni-Mn-Sn precipitates and show that \isotope[119]Sn-Möss\-bau\-er spectroscopy is an ideal technique to study the precipitate formation due to the possibility to probe nearest-neighbour interactions through the Sn nuclei. Therefore, we can observe phases with a short-range ordering otherwise absent or difficult to detect with XRD. We will show this trend by comparing our spectroscopic insights with X-ray diffraction results that resolve long-ranged ordering. \isotope[119]Sn-Mössbauer spectroscopy tracks the formation of stochiometric Ni$_\text{2}$MnSn clusters inside an antiferromagnetic Ni$_\text{50}$Mn$_\text{45}$Sn$_5$ matrix for mild annealing temperatures. Here, we indirectly probe the induced Sn-moment and use this spectroscopic feature as a detector for the structural transition, without the need for another tracing dopant used for example in Ref. \onlinecite{umetsu2008}, which leads to local distortions and effects the physical properties. On the other hand, X-ray diffraction is a well-known and effective tool to probe the long-range ordering of the whole sample volume. 
	
	\section{Results \& Discussion}
	
	\begin{figure*}[t!]
		\centering
		\includegraphics[width=.95\linewidth]{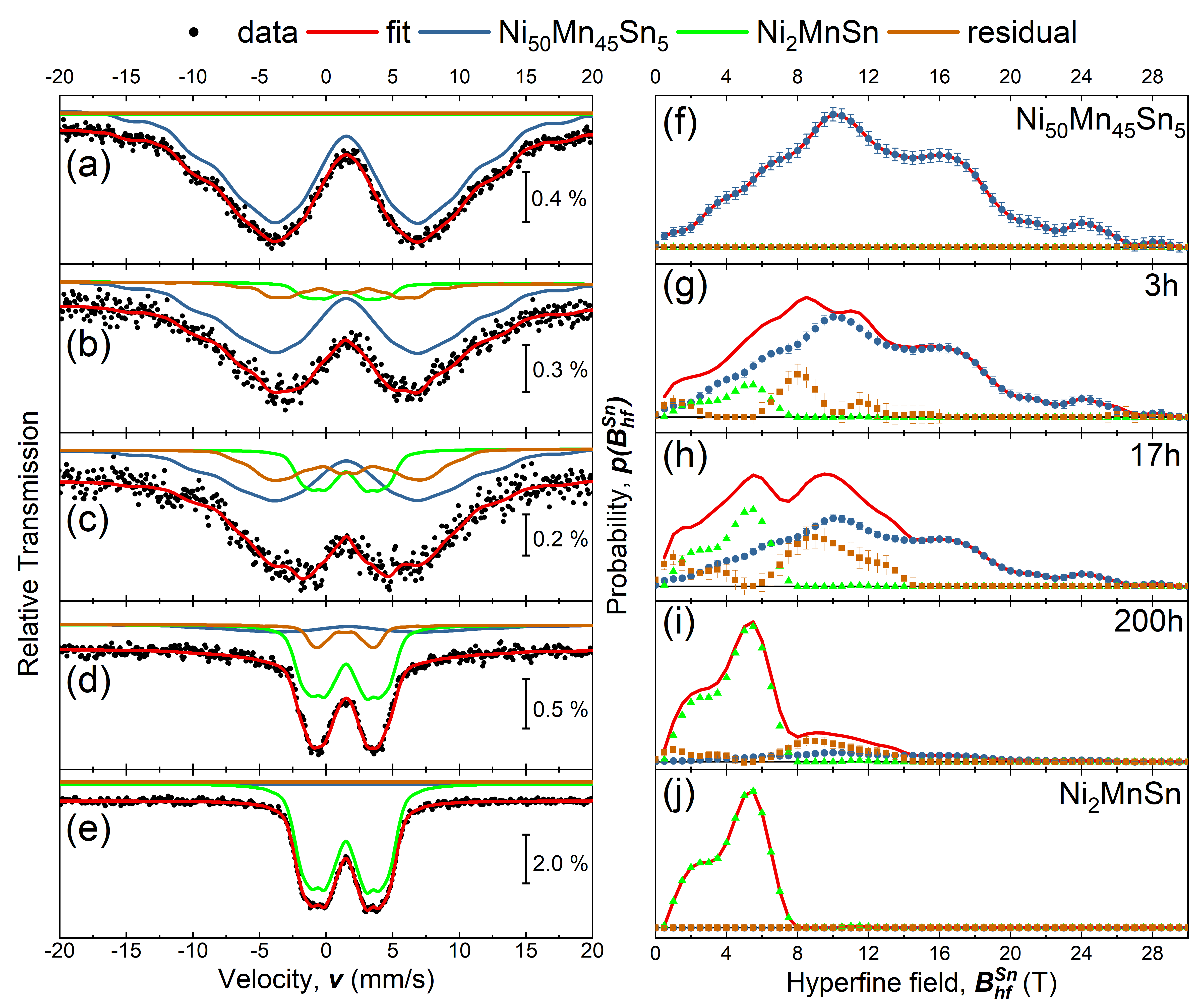}
		\caption{(a)-(e) Zero field Sn-M\"ossbauer spectroscopy measurements of the as-prepared Ni$_{50}$Mn$_{45}$Sn$_{5}$, Ni$_{50}$Mn$_{45}$Sn$_{5}$ heated at 650\,K for $t_A=3\mathrm{\,h}, 17\mathrm{\,h}, 200\mathrm{\,h}$ and for comparison an as-prepared stoichiometric Ni$_{2}$MnSn Heusler. All measurements were performed at room temperature. The individual spectra of the annealed states can be described by a linear combination of the spectra from the precursor materials (Ni$_{50}$Mn$_{45}$Sn$_{5}$ \& Ni$_{2}$MnSn) an additional residual distribution of hyperfine fields $p\left(B_{\text{hf}}^{\text{Sn}}\right)$. Additional details concerning the analysis of the M\"ossbauer spectra are discussed in the text. (f)-(j) Combined hyperfine field distribution $p\left(B_{\text{hf}}^{\text{Sn}}\right)$ for the respective measurements (red line), obtained by a combination of the sub spectra (blue, green and brown dots).}
		\label{fig:RT-Mossbauer}
	\end{figure*}
	In order to test the sensitivity of the Sn hyperfine field to structural changes, we performed \isotope[119]Sn-Mössbauer spectroscopy at room temperature on the AFM-ordered off stoichiometric L1$_0$-alloy Ni$_\text{50}$Mn$_\text{45}$Sn$_\text{5}$ and stoichiometric FM-ordered L2$_1$ Ni$_2$MnSn alloy (see Fig. \ref{fig:RT-Mossbauer}(a)\,\&\,(e))~\cite{Krenke2005}. The magnetic ordering of the samples leads to the lifting of the degenerated \isotope[119]Sn hyperfine levels and the occurrence of a sextet structure. In the following both spectra are described by a hyperfine field distribution $p\left(B_{\text{hf}}^{\text{Sn}}\right)$ which is shown in Fig. \ref{fig:RT-Mossbauer}(f)\,\&\,(j)) for the L1$_0$ and L2$_1$-ordered alloy, respectively. The spectrum of the L1$_0$-ordered alloy (see Fig. \ref{fig:RT-Mossbauer}(a)) possesses a relatively broad hyperfine field distribution ranging from 0\,T to almost 28\,T (see Fig. \ref{fig:RT-Mossbauer}(f)) with an average hyperfine field $\langle B_{\text{hf}}^{\text{Sn}}\rangle$ of  12.1\,T. The broad distribution of hyperfine fields indicates different local surroundings around the \isotope[119]Sn-nuclei. In contrast, the L2$_1$-ordered alloy feature a smaller distribution with distinct fine structure at 2 and 6\,T  leading to an average hyperfine field $\langle B_{\text{hf}}^{\text{Sn}}\rangle$ of 3.9\,T. This distribution of hyperfine fields indicates the presence of structural or magnetic disorder and expresses small variations from the L2$_1$-crystal structure. This phenomenon has been extensively discussed in a recent work on the effect of magnetic and anti-site disorder for the Sn-partial phonon density of states in Ni$_2$MnSn~\cite{Ni2MnSn}. These defects could be present in the form of anti-site disorder between Mn and Sn atoms caused due to the slight deviation of the perfect 2-1-1 stoichiometry (see Table\ref{tab:EDX-comp}). By comparing these two compositions, we can conclude that replacing Mn with Sn dilutes the absolute magnetic moment and, therefore, reduces the \isotope[119]Sn hyperfine field. Accordingly, the major contribution in the hyperfine field distribution (see Fig. \ref{fig:RT-Mossbauer}(f)\,\&\,(j)) shifts towards smaller fields and reflects the decreased Sn moment. Here, we cannot determine the exact change of the Sn magnetic moment due to the complex relationship between the magnetic moment and the hyperfine field. For Sn, the proportionality constant $A$ between the Sn moment $\mu_{\text{Sn}}$ and the Sn hyperfine field $B_{\text{hf}}^{\text{Sn}}$ depends on different materials properties, e.g. the anisotropy of the system~\cite{Delyagin1981,Mibu2000,Delyagin2003}. Furthermore, the effective hyperfine field measured at the \isotope[119]Sn nucleus is composed of several terms. There are direct (dipolar) and indirect (transfered hyperfine field from neighboring atom to \isotope[119]Sn nucleus) terms. There is also the possibility that the Sn atom is itself polarized from its surroundings and generates a direct (contact) hyperfine field. Without further information (for example XMCD spectra of Sn), we can only conclude that the measured $B_{\text{hf}}^{\text{Sn}}$ is the sum of relevant contributions. Therefore, the determination of the magnetic Sn-moment $\mu_{\text{Sn}}$ is beyond the scope of this work.
	
	In the following, the decomposition process will be investigated. As stated in previous investigations~\cite{Krenke2016,akr2016,Dincklage2018}, the decomposition process in the off-stoich\-i\-ome\-tric Heusler compound follows the route
	\begin{equation}
		\begin{split}
			5\,\text{Ni}_{50}\text{Mn}_{45}\text{Sn}_{5} &\longrightarrow \text{Ni}_{50}\text{Mn}_{25}\text{Sn}_{25} + 4\,\text{Ni}_{50}\text{Mn}_{50} \\
			&\longrightarrow \text{Ni}_{2}\text{Mn}_{}\text{Sn}_{} + 4\,\text{Ni}_{}\text{Mn}_{}.
		\end{split} 
		\label{eq:Decomp}
	\end{equation}
	In the following, we assume that for finite annealing times, the overall decomposition process can be described with an additional residual component, leading to a modification of Equation (\ref{eq:Decomp}) to
	\begin{equation}
		\begin{split}
			\text{Ni}_{50}\text{Mn}_{45}\text{Sn}_{5} &\longrightarrow \left(1-x-y-z\right)\cdot\text{Ni}_{50}\text{Mn}_{45}\text{Sn}_{5}   \\
			&+x\cdot \text{Ni}_{2}\text{Mn}_{}\text{Sn}_{} + y\cdot \text{Ni}_{}\text{Mn}_{} \\
			&+ z\cdot \sum_{\rho,\kappa,\epsilon} p_{\rho,\kappa,\epsilon} \cdot\text{Ni}_{\rho}\text{Mn}_{\kappa}\text{Sn}_{\epsilon},    
		\end{split}
		\label{eq:Decomp-Detail}
	\end{equation}
	where $\text{Ni}_{\rho}\text{Mn}_{\kappa}\text{Sn}_{\epsilon}$ is the residual Sn-containing phase with unknown stoichiometry, while x, y, z, and $p_{\rho,\kappa,\epsilon}$ is the respective volume fraction of the respective phase. 
	
	Due to excitation of the nuclear resonance, \isotope[119]Sn-M\"ossbauer spectroscopy probes only Sn-containing phases. Therefore, in the discussed case, one can track the temporal evolution of the decomposition process by identifying spectral fingerprints for the respective Sn-containing phases. Based on Eq. \ref{eq:Decomp-Detail}, it is possible to probe the initial compound Ni$_\text{50}$Mn$_\text{45}$Sn$_\text{5}$, the Ni$_2$MnSn-core structure, and the Sn-containing residual phase $\text{Ni}_{\rho}\text{Mn}_{\kappa}\text{Sn}_{\epsilon}$, while the formation of Ni-Mn can not be observed via \isotope[119]Sn-M\"ossbauer spectroscopy, due to the missing Sn-content. With the spectral fingerprint of the initial and the core-precipitate compound (see Fig.\ref{fig:RT-Mossbauer}(a)\,\&\,(e)) the experimental spectra of the decomposed state can be described by a least-squares fitting routine, assuming a superposition of the known theoretical models (for Ni$_{50}$Mn$_{45}$Sn$_{5}$ and Ni$_2$MnSn), while an additional hyperfine field distribution $p\left(B_{\text{hf}}^{\text{Sn}}\right)$ describes the residual spectral contributions arising from unknown compositions. Based on this model, we can model the \isotope[119]Sn M\"ossbauer spectra for annealing times $t_{\text{A}}$ of 3\,h, 17\,h, and 200\,h (see Fig. \ref{fig:RT-Mossbauer}(b)\,-\,(e)), while Fig. \ref{fig:RT-Mossbauer}(g)\,-\,(i) depicts the obtained hyperfine field distributions and the overall sum. Here the annealing temperature $T_{\text{A}}$ was chosen to be 650\,K, since at these temperatures similar studies on $\text{Ni}_{50}\text{Mn}_{45}\text{In}_{5}$~\cite{Dincklage2018} or $\text{Ni}_{50}\text{Mn}_{45}\text{Sb}_{5}$~\cite{Wanjiku2019} indicate that the size of the precipitate is almost temperature independent and below 10\,nm -- resulting in a sizeable surface-to-volume ratio. The size of the precipitates leads to the intrinsic compensation of the respective magnetic moments in the Ni-Mn shell and Ni$_2$MnSn core -- resulting in the shell-ferromagnetic effect~\cite{akr2016} and the occurrence of these large coercivity fields~\cite{Scheibel2017Room-temperatureShell-ferromagnet}. The comparison of the hyperfine field distributions reveals the shift of the maximum hyperfine field towards smaller values with increasing annealing times $t_{\text{A}}$, while after annealing for 200\,h, the majority of the hyperfine field contribution originates from stoichiometric Ni$_2$MnSn. The relative spectral area (see Table \ref{tab:SnMossbauerNiMnSn}) supports this trend: with increasing annealing duration, the Ni$_2$MnSn and residual contribution increase. 
	
	\begin{table}[h]
		\centering
		\caption{Relative spectral area obtained from the hyperfine field distribution $p\left(B_{\mathrm{hf}}\right)$ for the different components.}
		\begin{tabular}{l|c|c|c}
			$t_A$ 	& $\text{Ni}_{50}\text{Mn}_{45}\text{Sn}_{5}$		&	$\text{Ni}_2\text{MnSn}$		&	residual \\
			(h) 	& (\%)			&	(\%)		&	(\%) \\ \hline \hline
			0		&	100			&	0			&	0	\\
			3		&	80.8(1.8)	&	11.3(1.1)	&	7.9(9)	\\
			17		&	57.9(2.3)	&	19.8(2.1)	&	22.3(1.6)	\\
			200		&	15.2(1.5)	&	66.8(1.7)	&	18.0(1.1)	\\
		\end{tabular}
		\label{tab:SnMossbauerNiMnSn}
	\end{table}
	\begin{figure}[h!]
		\centering
		\includegraphics[width=.7\linewidth]{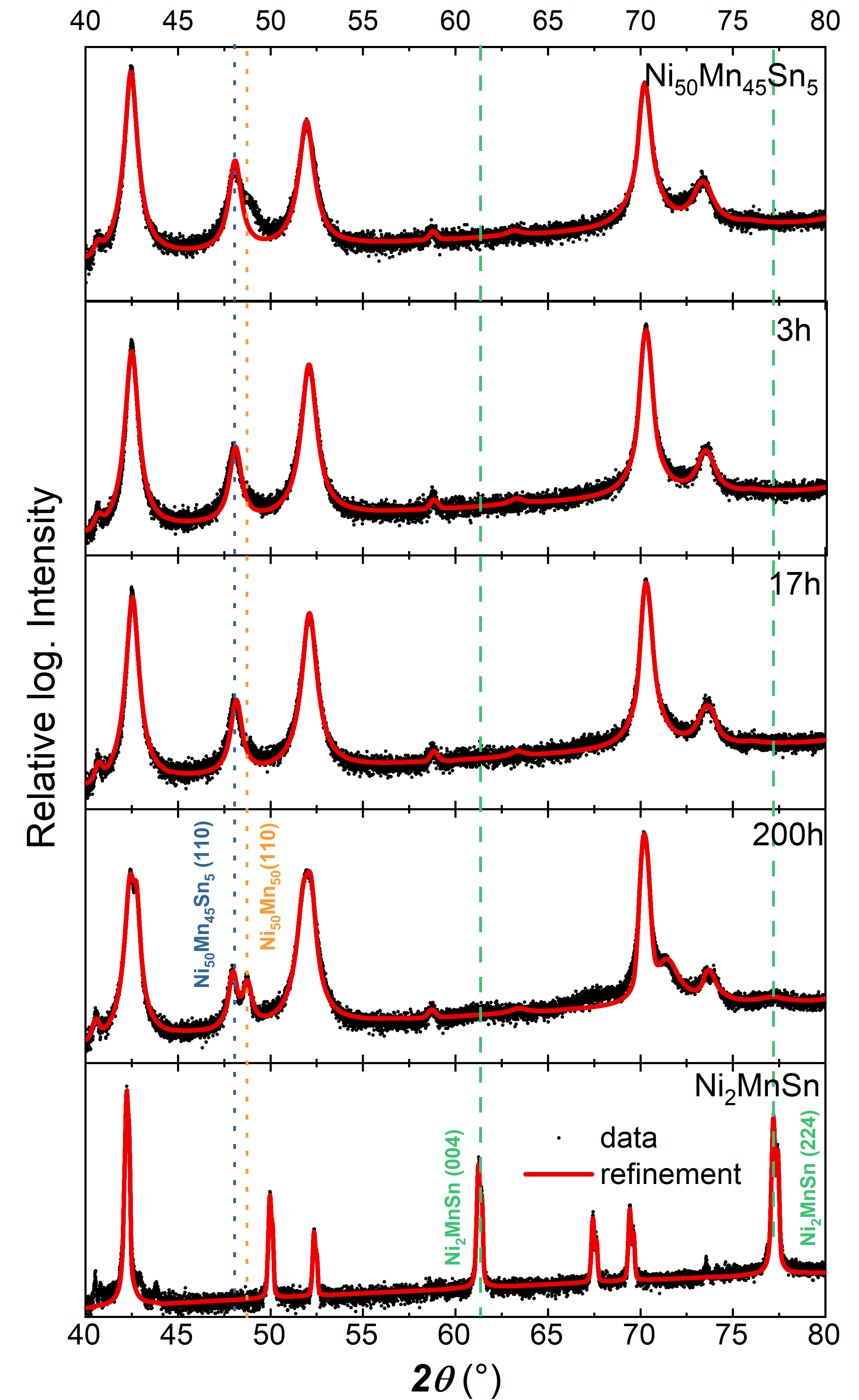}
		\caption{Comparison of XRD diffractograms (black circles) for the different annealed states and the corresponding refinements (red line). Certain bragg peaks that can be used as a finger print of the respective phase have been labelled.}
		\label{fig:XRD-Stack}
	\end{figure}	
	X-ray diffraction probes the long-range ordering of the lattice structure. Figure \ref{fig:XRD-Stack} depicts the X-ray diffractograms for the decomposed states after 3, 17, and 200\,h for annealing temperature $T_A$ of 650\,K. These diffractograms indicate that $\text{Ni}_{50}\text{Mn}_{45}\text{Sn}_{5}$ crystallizes in its initial  L1$_0$-phase, similar to Ni-Mn with a small deviation of the lattice constant, while the stoichiometric compound Ni$_2$MnSn possess a L2$_1$ ordering. Only after annealing the sample for 200\,h, the decomposition becomes visible in the XRD-pattern as a splitting of the (110) L1$_0$-peaks. Additional detailed analysis shows that the (004) and (224) peaks of the L2$_1$ full Heusler Ni$_\text{2}$MnSn alloy are barely visible after annealing for 200\,h. The small contribution of the L2$_1$-phase in the XRD pattern can be explained by the relatively small precipitate size~\cite{Dincklage2018,Wanjiku2019} (below 10\,nm) and the low phase fraction of the formed full-Heusler precipitates. For the investigated post-annealing conditions, the XRD patterns indicate that the long-range ordering of the sample has barely changed, while M\"ossbauer spectroscopy reveals drastic variations of the Sn nearest neighbour surrounding. These variations of the local surrounding is reflected in the nuclear hyperfine levels.
	
	\section{Summary}
	Within this work, we investigated the formation of Ni$_\text{2}$MnSn-precipitates in off-stoichiometric $\text{Ni}_{50}\text{Mn}_{45}\text{Sn}_{5}$ via X-ray diffraction and M\"ossbauer spectroscopy. Here, the transferred hyperfine field (or the induced magnetic moment) of \isotope[119]Sn is an interesting property for investigating and tracking the precipitation process. While X-ray diffraction reveals long-range ordering, \isotope[119]Sn-M\"oss\-bauer spectroscopy probes nearest-neighbour interactions and is, therefore, especially sensitive to changes in the local surrounding. Due to these differences in the probed length scale, we can explain the different occurring dynamics. Employing extended X-ray absorption fine structure (EXAFS) spectroscopy, this concept can be adapted to different material systems. For example, one can probe the diffusion at grain boundaries~\cite{Liu2021} in Nd$_2$Fe$_{14}$B or Sm-Co~\cite{Swilem1999}, or one can use scanning transmission X-ray microscopy~\cite{Suzuki2016} and find a connection between the local structure and high coercivity occurring in high-performance permanent magnets.
	
	\section{Acknowledgements}
	This~work was funded by the Deutsche Forschungsgemeinschaft (DFG, German Research Foundation) within TRR~270 (subprojects~A03, A04, B01, and B05), Project-ID~405553726. 
	
	\section{Experimental Details}
	
	All samples were prepared by arc melting of pure elements (Ni: 99.98\,\%, Mn: 99.99\,\%, Sn: 99.999\,\%). Afterwards, the obtained material was encapsulated in a quartz tube under argon atmosphere and homogenized for five days at 1073\,K , followed by quenching in room temperature water and polished. Energy-dispersive X-ray spectroscopy inside a scanning electron microscope verifies the composition of the prepared alloys (see Table \ref{tab:EDX-comp}). For the investigation of the decomposition, Ni$_{\text{50}}$Mn$_{\text{45}}$Sn$_{\text{5}}$ was annealed a temperature of 650\,K for different times under high vacuum conditions ($p\approx 5\cdot10^{-5}\text{\,mbar}$) to avoid oxidation of the sample. Room temperature \isotope[119]Sn-Mössbauer spectroscopy measurements were performed in transmission geometry under zero-field conditions with conventional electronics. The velocity of a Ca\isotope[119]SnO$_\text{3}$ source was changed within the constant-acceleration mode and calibrated with a laser interferometer. The experimental spectra have been evaluated by a least-squares fitting routine using the $Pi$ program package~\cite{PiLink}. X-ray diffraction measurements were performed using a Phillips PANalytical X'Pert PRO with non-monochromatized X-rays (Cu X-ray source) in a Bragg-Brentano geometry, and the obtained diffraction patterns were analyzed using JANA2006~\cite{Jana}.
	
	\begin{table}[h]
		\centering
		\caption{Composition of the different samples determined by EDX analysis.}
		\begin{tabular}{c|c|c|c}
			&   Ni  &   Mn  &   Sn\\ \hline \hline
			Ni$_\text{50}$Mn$_\text{45}$Sn$_\text{5}$      &   50.3  &   44.7  &   5.0\\
			Ni-Mn            &   49.6  &   50.4  & \\
			Ni$_2$MnSn      &   49.2  &   24.2  &   26.6\\
		\end{tabular}
		\label{tab:EDX-comp}
	\end{table}
	
	
	%

\end{document}